\documentclass[manuscript=article]{achemso}

\usepackage{graphicx}
\usepackage{amsmath}
\usepackage{bm}
\usepackage{textcomp}
\usepackage[english]{babel}
\usepackage[utf8]{inputenc}
\usepackage{times}
\usepackage{cancel}
\usepackage[T1]{fontenc}
\usepackage{color}
\usepackage{here}
\usepackage{amsfonts}
\usepackage[percent]{overpic}
\usepackage{chemformula}
\usepackage[version=4]{mhchem}
\usepackage{hyperref}
\usepackage{mathpazo}
\usepackage{etoolbox}
\usepackage{xr}
\usepackage{multirow}

\title{Current rectification by nanoparticles in bipolar nanopores}

\author{Andr\'{e}s~C\'{o}rdoba}
\affiliation{Pritzker School of Molecular
Engineering, University of Chicago, Chicago, Illinois 60637, United States}
\altaffiliation{Advanced Materials for Energy-Water Systems
(AMEWS) Energy Frontier Research Center, Argonne
National Laboratory, Lemont, Illinois 60439, United States}
\author{Joan M. Montes de Oca}
\affiliation{Pritzker School of Molecular
Engineering, University of Chicago, Chicago, Illinois 60637, United States}
\altaffiliation{Advanced Materials for Energy-Water Systems
(AMEWS) Energy Frontier Research Center, Argonne
National Laboratory, Lemont, Illinois 60439, United States}
\author{Johnson Dhanasekaran}
\affiliation{Pritzker School of Molecular
Engineering, University of Chicago, Chicago, Illinois 60637, United States}
\altaffiliation{Advanced Materials for Energy-Water Systems
(AMEWS) Energy Frontier Research Center, Argonne
National Laboratory, Lemont, Illinois 60439, United States}
\author{Seth B. Darling}
\affiliation{Pritzker School of Molecular
Engineering, University of Chicago, Chicago, Illinois 60637, United States}
\alsoaffiliation{Chemical Sciences and Engineering Division, Argonne
National Laboratory, Lemont, Illinois 60439, United States}
\altaffiliation{Advanced Materials for Energy-Water Systems
(AMEWS) Energy Frontier Research Center, Argonne
National Laboratory, Lemont, Illinois 60439, United States}
\author{Juan~J.~de Pablo}
\affiliation{Pritzker School of Molecular
Engineering, University of Chicago, Chicago, Illinois 60637, United States}
\alsoaffiliation{Materials Science Division, Argonne National Laboratory,
Lemont, Illinois 60439, United States}
\altaffiliation{Advanced Materials for Energy-Water Systems
(AMEWS) Energy Frontier Research Center, Argonne
National Laboratory, Lemont, Illinois 60439, United States}
\email{depablo@uchicago.edu}

\begin{document}

\begin{abstract}

Bipolar nanochannels comprising two domains of positively and negatively charged walls along the pore axis 
are known to rectify current when exposed to an \textcolor{black}{electric potential bias}. 
We find that addition of charged nanoparticles can increase rectification considerably, by approximately one order of magnitude. Two bipolar channel geometries are considered here; their behavior is examined at rest and under the influence of a negative bias and a positive bias, respectively. We do so by relying on a molecular-level model of the electrolyte solution in the channels.
The large increase in current rectification can be explained by the 
inherent electric field that charged nanoparticles generate within the channel. This effect is found to be largely dependent on 
the pore's geometry, its charge distribution, and the sign of the nanoparticles' charge, thereby offering new opportunities for design of engineered nanopore membrane-nanoparticle systems for energy storage.
\end{abstract}

\section{Keywords}

\noindent Bipolar nanopores, molecular dynamics, nanoparticles, 
electric current rectification, power generation

\section{Introduction}

Permselective-media-based bipolar diodes consist of pores with at least 
two regions of opposite charge. 
These bipolar nanopores are interesting from both a fundamental point of view and from a technological perspective. 
Bipolar nanopores can exhibit powerful rectification properties due to the 
asymmetry of a charge pattern along
the wall of the pore. In this context, rectification refers to the nanpores' ability to 
generate a preferred direction in the transport of electric current.
For the particular case of bipolar pores having separate positive 
and negative surface charge domains along the pore axis, 
rectification is significant if the radius of the nanopore is small compared to the
screening length of the electrolyte. This ionic rectification is accompanied by the formation of depletion zones for both cations and anions along the main axis of the pore, whose extent depends on the
magnitude and sign of the applied external voltage \cite{fertig2020rectification}.
A bipolar diode comprises a negatively \textcolor{black}{charged} region and a positively charged region; 
when their surface charge densities have the same absolute value, one has a so-called nanofluidic diode. 
By modifying the surface charge density in the middle section of the channel, one can form a 
nanofluidic bipolar transistor \cite{daiguji2005nanofluidic}.
Experimental work on bipolar nanopores has examined a range of properties, including 
ionic current rectification, breakdown, and switching in heterogeneous 
oxide nanofluidic devices \cite{cheng2009ionic, vlassiouk2007nanofluidic}.

\textcolor{black}{Janus membranes composed of bipolar nanopores 
have recently been used to increase the efficiency of 
reverse electrodialysis (RED). In RED, an electric current is
harvested due to the asymmetric concentration of ions that arises
in an electrolyte when held in two reservoirs
connected by a membrane system \cite{post2008energy, brogioli2009extracting}. 
At large scales this can be accomplished by placing RED membranes at 
river mouths where river water mixes with sea water.
In the typical membrane-based RED process, the practical power output
is limited by non-Ohmic mass-transfer resistances. These resistances are
caused by the accumulation of ions inside the membrane's charged nanopores
near the side that interfaces with the low concentration reservoir 
\cite{post2008energy,chinaryan2014effect}.
This accumulation lowers the concentration difference across the membrane and
suppresses ionic transport. This is a common limitation in traditional RED but it
can be eliminated with an asymmetric bipolar membrane, which
considerably increases the output power density 
\cite{gao2014high,zhang2015engineered,zhang2017ultrathin,zhu2018unique}. 
In this work we do not impose a concentration difference across the 
nanopores, we instead apply a voltage bias, 
and therefore we do not explicitly model a RED process. However, with the approach
employed here the effect of small concentrations of charged nanoparticles in 
the current-voltage relationship of bipolar nanopores can be examined.}

While important theoretical and experimental advances have led to a better understanding of 
the fundamental physics at work in bipolar nanopores, the coupled roles of 
nanopore geometry and the addition 
of small concentrations of charged nanoparticles on ion transport
and electric current rectification have not been considered before.
In previous work \cite{deOca_dePablo_2022}, we calculated and analyzed the cation and anion
concentration profiles and current-voltage
relationship of dilute electrolytes in bipolar nanopores
at rest and under applied biases of $+2~\text{V}$ and $-2~\text{V}$, respectively.
In this work we discuss the effect of adding small quantities of nanoparticles
on the concentration profiles and current 
rectification properties of bipolar nanopores. Our results indicate that rectification can be increased 
considerably, raising distinct opportunities for development of practical devices for energy storage.

Several theoretical studies of the current-voltage 
relationship have been reported for membranes consisting of two fixed 
charge regions of opposite sign. 
That body of work has relied on solution of the underlying diffusion equations in
conjunction with the Poisson-Boltzmann equation
\cite{mauro1962space, coster1965quantitative}.
An analysis of 
four 1D permselective membranes adjacent to each other has 
also been presented \cite{sonin1972ion}. Due
to the complexity of such systems, however, a number of
simplifications have generally been made, 
such as assuming linear concentration and electric potential profiles within the permselective regions \cite{sonin1972ion}. 
A simple theory for multi-ionic transport, non-equilibrium water dissociation, 
and space-charge effects in bipolar membranes was developed 
on the basis of some of the concepts used to describe solid-state
P-N junctions \cite{mafe1990model}; specifically, 
ion transport was modeled in terms of the Nernst-Planck flux 
equation and non-equilibrium water dissociation was accounted for 
by the Onsager theory of the second Wien effect \cite{mafe1990model}. 
An important finding of that study was the concept of rectification, 
whereby a preferred direction in the transport of current can be achieved.

Analytical expressions for the 
current-voltage, or I-V response of bipolar nanopores have been derived 
under a set of \textit{ad hoc} assumptions. With these 
expressions \cite{green2015asymmetry}, it has been proposed that, in theory, the surface charge 
asymmetry could lead to simple nanofluidic-based diodes whose rectification
(depending on geometry) can be as large as $10^2-10^3$.
This means that the electric current generated by, for example, 
a positive voltage can be up to a thousand times larger than the 
current generated by a voltage of the same magnitude but negative 
sign. Other work has investigated the effects of a single 
non-ideal permselective region 
(three-layers system) \cite{chinaryan2014effect}, and a complete theoretical model of a bipolar diode
consisting of a four-layered model (two microchambers and
two permselective regions) has also been derived
under the assumption of cross-sectional local electroneutrality
\cite{green2015asymmetry}.
More recently, the possibility of nano-fluidic RED
 for energy harvesting from salinity gradients has been evaluated by
considering ion transport in bilayer cylindrical nanochannels
consisting of different sized nanopores, connected to two large reservoirs having 
different NaCl concentrations. In that work, numerical
simulations at the level of coupled Poisson-Nernst-Planck and Navier-Stokes
equations have been used to describe the electrokinetic behavior over 
asymmetric sub-pore lengths, and to predict the effects of surface charge
on transference number, 
osmotic current, diffusive voltage,
maximum power and maximum power efficiency \cite{long2018reverse}. 

Multiscale modeling approaches have also 
been used before to study charged
nanopores \cite{moy2000tests, corry2000tests, valisko2019multiscale}.
The validity of the mean-field approximation in Poisson-Nernst-Planck theory 
has been established by contrasting its predictions with those of Brownian dynamics simulations in 
both model cylindrical channels and in a more realistic potassium
channel. However, in simple cylindrical channels,
considerable differences arise between these two levels of description with regards to the 
concentration profiles in the channel and the corresponding conductance properties. 
These differences are more pronounced
in narrow channels having a radius smaller than the Debye length,
and gradually diminish with increasing radius. Convergence occurs when the channel 
radius is over two Debye lengths \cite{moy2000tests,corry2000tests}. 

In recent years, the ability to 
fabricate carefully designed nanopores has renewed interest in such systems. Bipolar nanochannels have been studied 
using three modeling levels \cite{fertig2020rectification, valisko2019multiscale}  
that include (1) an all-atom explicit-water 
model studied with molecular dynamics, and reduced models with 
implicit water containing (2) hard-sphere ions studied through a ``Local
Equilibrium Monte Carlo'' simulation method 
(which determines ionic correlations accurately), and (3) point ions studied with 
Poisson-Nernst-Planck theory (mean-field approximation). 
It has been shown that reduced models are able to 
reproduce key device functionalities (\textit{e.g.} rectification and selectivity)
for a wide variety of charge patterns; that is, reduced 
models are useful in understanding the mesoscale physics 
of the device (\textit{e.g.} how the
current is produced). Multiscale approaches 
have also been used to examine the relationship of  
reduced implicit-water models to explicit-water models, and have proposed 
that diffusion coefficients in reduced models can be treated as adjustable 
parameters for comparisons between explicit- and implicit-water
models \cite{valisko2019multiscale}. 
It has also been reported that the values of the diffusion coefficients 
are sensitive to the net charge of the pore, but are relatively
transferable to different voltages and charge patterns with the same total charge
\cite{valisko2019multiscale}.

\textcolor{black}{In this work we employ molecular dynamics simulations 
with implicit ions and an implicit solvent. The parameters for the ion-ion  effective potentials
are obtained from systematically coarse-graining all-atom interactions.
The use of an implicit solvent 
does not allow to account for electroconvective
mechanisms in the ionic transport inside the bipolar nanopores 
\cite{rubinstein2000electro,zaltzman2007electro,gubbiotti2022electroosmosis}.
However in charged nanopores there is a regime where a decoupling 
of the electrodiffusive problem from the electroconvective
problem occurs. This decoupling is determined by the smallness
of the electroconvective P{\'e}clet number 
\cite{rubinstein1991electroconvection,green2014effects}. In bipolar nanopores
the assumption of $\text{P{\'e}}\ll1$ holds in the low-voltage
Ohmic region of the I-V curve, at voltages beyond the Ohmic region
convection effects can become important. Models of charged nanopores
based on the Poisson-Nernst-Planck theory 
\cite{green2014effects,green2015asymmetry,chinaryan2014effect,green2015time} 
are based on the $\text{P{\'e}}\ll1$ assumption.
The MD simulations presented here are valid in that same regime. Moreover 
the discrete nature of the simulation method allows us to study the effect that
nanoparticles with a surface charge distribution have on the rectification properties 
of bipolar nanopores.}

\section{Results and Discussion}

\subsection{Effect of nanoparticle addition on the concentration distributions}

Two bipolar nanopore geometries are examined in what follows, namely
a pore in which the two oppositely charged sections have the same length
and wall charge density, and another in which the section with positive wall 
charges is four times shorter and has a wall charge density three times larger than the 
section with negative wall charges. In the first nanopore, referred to as C1, 
the section with positively charged walls has a charge density of
$\sigma_1=+0.1~C/{\rm m}^2$. The section with negatively charged walls has a wall charge density of
$\sigma_2=-0.1~C/{\rm m}^2$. The wall charge density is 
$\sigma_{\neq 1,2}=0~C/{\rm m}^2$ for the rest of the walls.
In the second channel, which is referred to as C2,
the region with positively charged walls has a wall charge density of
$\sigma_1=+0.24~C/{\rm m}^2$, and 
the section with negatively charged walls has a wall charge density of
$\sigma_2=-0.08~C/{\rm m}^2$. In C2 the wall charge density is 
also $\sigma_{\neq 1,2}=0~C/{\rm m}^2$ for the rest of the walls. \textcolor{black}{
The magnitudes of the surfaces charges adopted here are comparable to 
those found in experimental studies with Janus membranes
\cite{gao2014high,zhang2015engineered,zhang2017ultrathin,zhu2018unique}.}
Both geometries feature changes in the height as a function 
of channel length. In both channels, 
the section with positively charged walls has a height of $17$ nm 
and the section with negatively charged walls has a height of $10$ nm.
\textcolor{black}{When nanoparticles are added to the nanopores, 
the ionic strength of the solution ($\approx0.1$ M)
is kept constant. The Debye length of the electrolytic solution is 
$\lambda_{\rm D}\approx 0.95~\text{nm}$. Which is an order of magnitude 
smaller than the height in the narrowest section of the nanochannels considered here.
Plots of the ionic concentrations as a function of the channel's height 
in the wide and narrow sections of the nanopore are given 
in Figure S1 in the Supporting Information (SI).}

We consider the addition of 10 nanoparticles in both channels.
Moreover the effect of changing the surface charge of the nanoparticles 
from positive to negative is also studied here. \textcolor{black}{Each nanoparticle has
a hydrodynamic radius of about 1.14 nm and a surface charge density of about
$\pm0.33~\text{C}/\text{m}^2$. See the Methods section below for more details 
on the nanoparticle model used here.}
The contributions to the ionic strength from the cations, anions and 
nanoparticles for the 
different cases considered here are summarized in Table \ref{tab:part}.

\begin{table}[ht]
\centering
\renewcommand{\arraystretch}{1.2}
\begin{tabular}{ccccccccc}
\hline
Additive & \multicolumn{2}{c}{$n_c z_c e$} & \multicolumn{2}{c}{$n_a z_a e$}
& \multicolumn{2}{c}{$n_w z_w e$} & \multicolumn{2}{c}{$n_p z_p e$} \\
\hline 
& C1 & C2 & C1 & C2 & C1 & C2 & C1 & C2\\
\hline
None & $+1535e$ & $+1609e$  & $-1535e$ & $-1409e$ & $0e$ & $-200e$  & $0e$ & $0e$\\ 
\hline
& \multicolumn{8}{c}{Negatively Charged}\\
\hline
10 Particles & $+1535e$ & $+1609e$  & $-1455e$ & $-1329e$ & $0e$ & $-200e$  & $-80e$ & $-80e$ \\
\hline
& \multicolumn{8}{c}{Positively Charged}\\
\hline
10 Particles & $+1455e$ & $+1529e$  & $-1535e$ & $-1409e$ & $0e$ & $-200e$  & $+80e$ & $+80e$ \\
\hline
\end{tabular}
\caption{Summary of the contribution of charges from each of the different components 
of the nanopores considered here. Note that for each channel, the ionic strength is kept 
constant when nanoparticles are added by removing a certain number of cations or anions.
$n_c z_c e$ is the total charge from cations, 
$n_a z_ae$ is the total charge from anions, 
$n_w z_we$ is the total charge from wall sites and $n_p z_pe$ is the total charge from
particle sites. }
\label{tab:part}
\end{table}

\begin{figure}[h t]
\includegraphics[width=\linewidth]{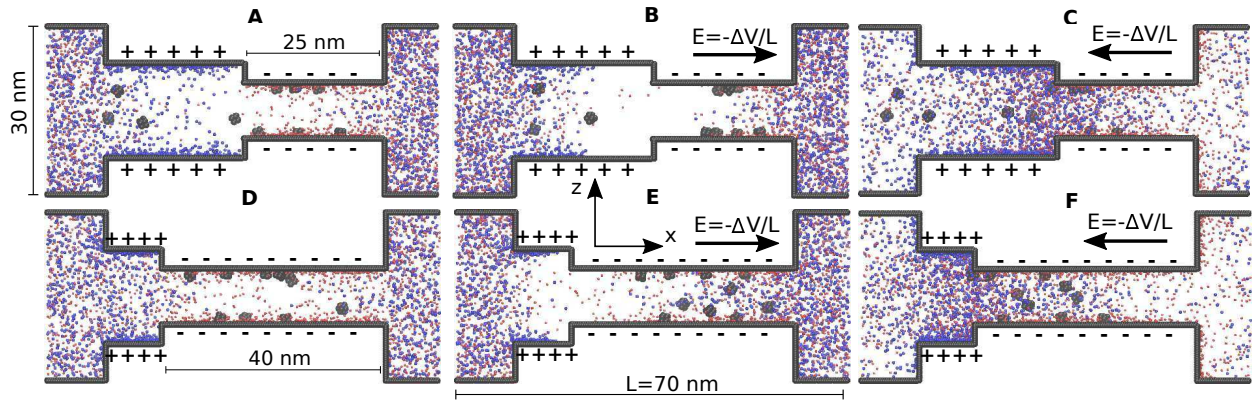}
\caption{Orthographic projections of steady-state configurations for the two bipolar 
slit nanopores considered in this work. The top row shows nanopore C1 and the 
bottom row nanopore C2.
Sodium ions are represented as
red spheres and chlorine ions as blue spheres. 
Black spheres represent the interaction sites that make up 
the nanoparticles and the channel walls. The cases shown correspond to
positively charged nanoparticles. The cases 
with negatively charged nanoparticles are shown in the 
Supporting Information file. In A and D no bias is applied (\textit{i.e.} equilibrium). B and E have a 
negative bias, $-2~\text{V}$, applied. And C and F have a
positive bias, $+2~\text{V}$, applied.}
\label{figRP}
\end{figure}

Figure \ref{figRP} shows steady state
orthographic projections of the two bipolar nanopores
as predicted by the coarse-grained molecular dynamics (MD) simulations. 
Sodium ions are represented as red spheres and Chlorine ions as blue spheres
and the black spheres represent the interaction sites that make up 
the channel walls and the nanoparticles. The cases shown in Figure \ref{figRP}
correspond to nanopores where positively charged nanoparticles have been added.
The same channels but with negatively charged nanoparticles can be found in
Figure S2 of the SI.
Figures \ref{figRP} A and D correspond to C1 and C2, respectively, when no voltage 
bias is applied.
It can be observed that there is an obvious excess of anions in the wider sections 
of the nanopores, where the wall has a positive charge. In the narrow
section of the nanopores, where the wall charge is negative, 
a excess of cations is evident. In C1 the nanoparticles are distributed 
more or less evenly between the wide and narrow sections of the pore.
In C2 the nanoparticles mostly remain in the long narrow section.
With respect to the height of the channels, the location of the nanoparticles 
does not exhibit major differences between C1 and C2.
The positively charged nanoparticles 
reside near the wall in the section of the pores with negative wall charge.
In the section of the pores with positive wall charge the positively charged
nanoparticles tend to stay near the center of the channels. As expected, the 
negatively charged nanoparticles do the opposite.

Figures \ref{figRP} B and E show C1 and C2, respectively, when a negative bias, $-2~\text{V}$, 
is applied. \textcolor{black}{The details on how the voltage
bias between the two ends of the nanopores is applied in the simulations are given 
in the Methods section below. Also note that the magnitude of the voltage bias used here 
is what has been commonly used in experiments with Janus membranes 
\cite{gao2014high,zhang2015engineered,zhang2017ultrathin,zhu2018unique}.} 
In this case, a pronounced depletion zone, where the amount of both anions 
and cations drops, is observed around the transition from the narrow to 
the wide section of the nanopores. For this negative bias the general location 
of the nanoparticles inside the channels is not significantly changed compared to the 
equilibrium cases. However, in C1 it is more apparent that, 
similarly to what the ions do,
the nanoparticles also stay away from the aforementioned depletion zone. 
Figures \ref{figRP} C and F show C1 and C2, respectively, 
when a positive bias $+2~\text{V}$, is applied. Here, 
contrary to what happens in the negative bias case, anions and cations
appear to concentrate around the transition zone between the narrow and 
wide sections of the channels. For the positive bias, 
the positively charged nanoparticles 
move towards the wide section of the nanopores. 
They tend to concentrate around the transition zone between the narrow 
and wide sections of the cahnnels.
As can be seen in Figure S3 in the SI, the negatively charged nanoparticles move towards the 
narrow section when the positive bias is applied.

\begin{figure}[h t]
\center
\includegraphics[width=\linewidth]{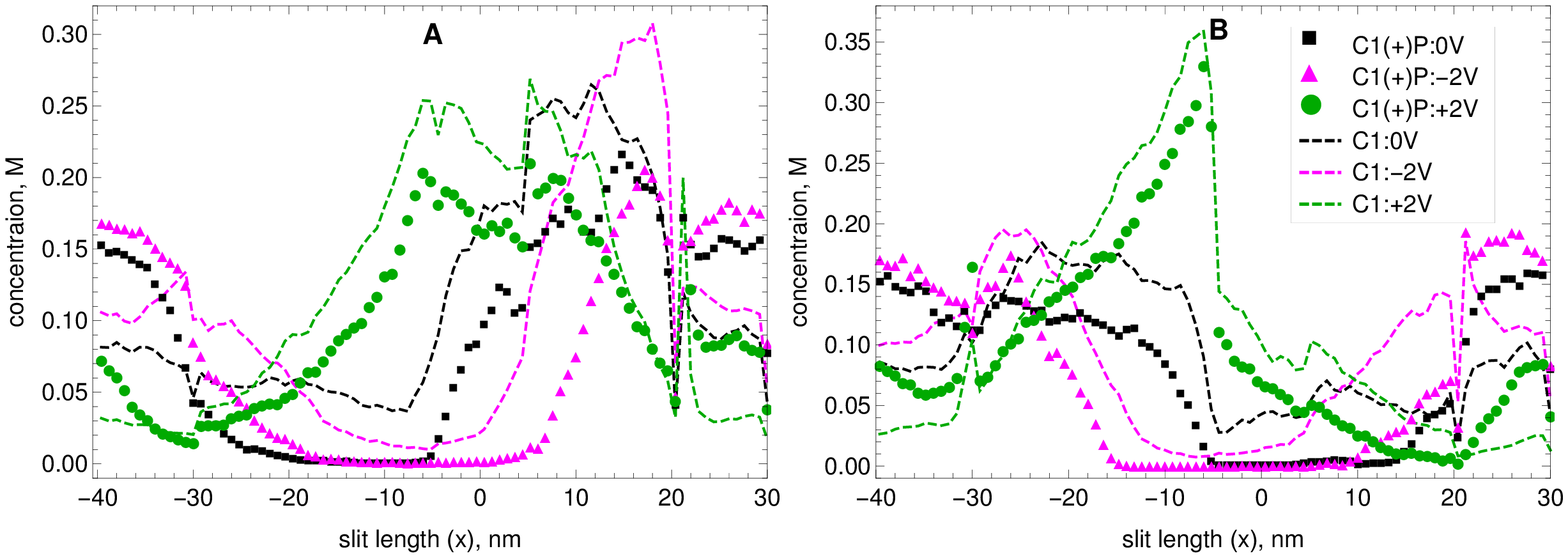}\\
\includegraphics[width=\linewidth]{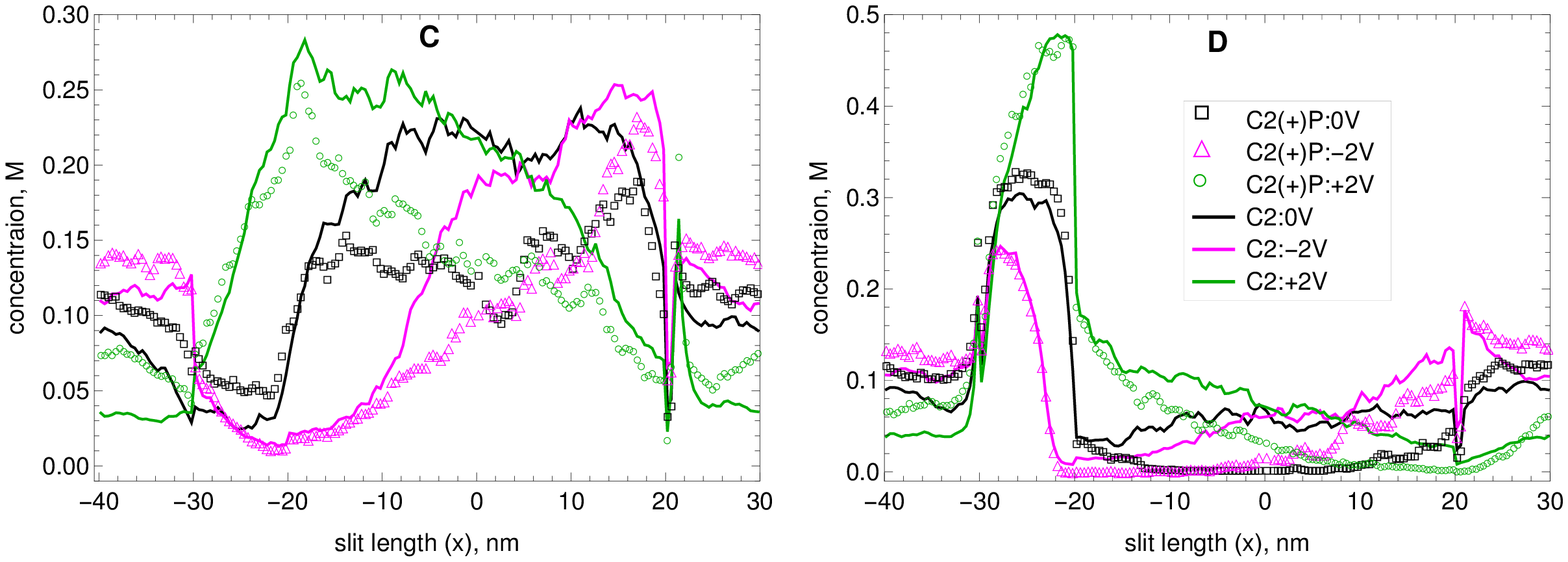}
\caption{Average concentration as a function of channel length
of A) cations and B) anions in C1 at equilibrium 
and for two different applied 
voltages with (symbols) and without (dashed lines) the addition 
of positively charged nanoparticles.
Concentration of C) cations  and D) anions in C2 at equilibrium 
and for two different applied 
voltages with (symbols) and without (lines) the addition of positively 
charged nanoparticles. The concentration profiles with addition of negatively 
charged nanoparticles are given in the Supporting Information file.}
\label{figconcpp}
\end{figure}

The top row of Figure \ref{figconcpp} shows concentration profiles
as a function of the channel's length \textcolor{black}{(\textit{i.e.} averaged over the channel's
width and depth)}. for (A) cations and (B) anions in C1 without 
nanoparticles (lines) and in C1 with positively charged nanoparticles (symbols). 
The bottom row of Figure \ref{figconcpp} shows concentration profiles
as a function of the channel's length
for (A) cations and (B) anions in C2 without 
nanoparticles and in C2 with positively charged nanoparticles (symbols).
The same plots but for the cases where the nanoparticles have a negative surface
charge are provided in the SI.
As expected, the concentration of cations is larger in the 
narrow section of the slits, where the walls are negatively charged.
In the wide section of the slits, where the wall is positively charged, 
the concentration of cations is smaller. A depletion zone is observed 
where the concentration of cations 
drops nearly to zero in the transition region
between the narrow and wide sections of the 
nanopores. This depletion zone is larger for the $-2~\text{V}$ bias
case, and is also present in the equilibrium case. However, it is almost non-existent 
in the $+2~\text{V}$ bias case.
It has been previously reported that both cations and anions exhibit
depletion zones in bipolar nanochannels, and that the depth of these depletion zones is sensitive to 
sign of the external voltage \cite{fertig2020rectification}. 
The presence of either the negatively or the positively charged nanoparticles increases the 
length and depth of this depletion zone for the cations in both channels. It is also evident
from Figure \ref{figconcpp} that the cation depletion zone is longer and deeper in C1. Furthermore, it is
the longest and deepest for C1 with a $-2~\text{V}$ applied bias. For the $+2~\text{V}$ 
applied bias no depletion zone is observed for the cations, and the maximum 
concentration occurs in the transition region
between the narrow section and the wide section of the 
nanopores.

As can be seen in Figures \ref{figconcpp} B and D, the concentration of anions is 
larger in the wide section of the slits, where the walls are positively charged.
In the narrow section of the slits, where the wall is negatively charged, 
the concentration of anions is smaller. A depletion zone is also observed 
where the concentration of anions
drops nearly to zero in the transition region
between the narrow section and the wide section of the 
pore. This depletion zone again is observed at equilibrium, 
but is most pronounced in the 
$-2~\text{V}$ case. As before, nanoparticles make the anion depletion zone 
significantly longer and deeper, specially in the C1 nanopore. 
The highest concentrations of anions occur for a $+2~\text{V}$
bias applied, and they concentrate near the transition area between the 
narrow and wide sections of the slit. Note that they stay primarily on the 
side where the wall is positively charged.

\begin{figure}[h t]
\center
\begin{overpic}[width=0.48\linewidth]{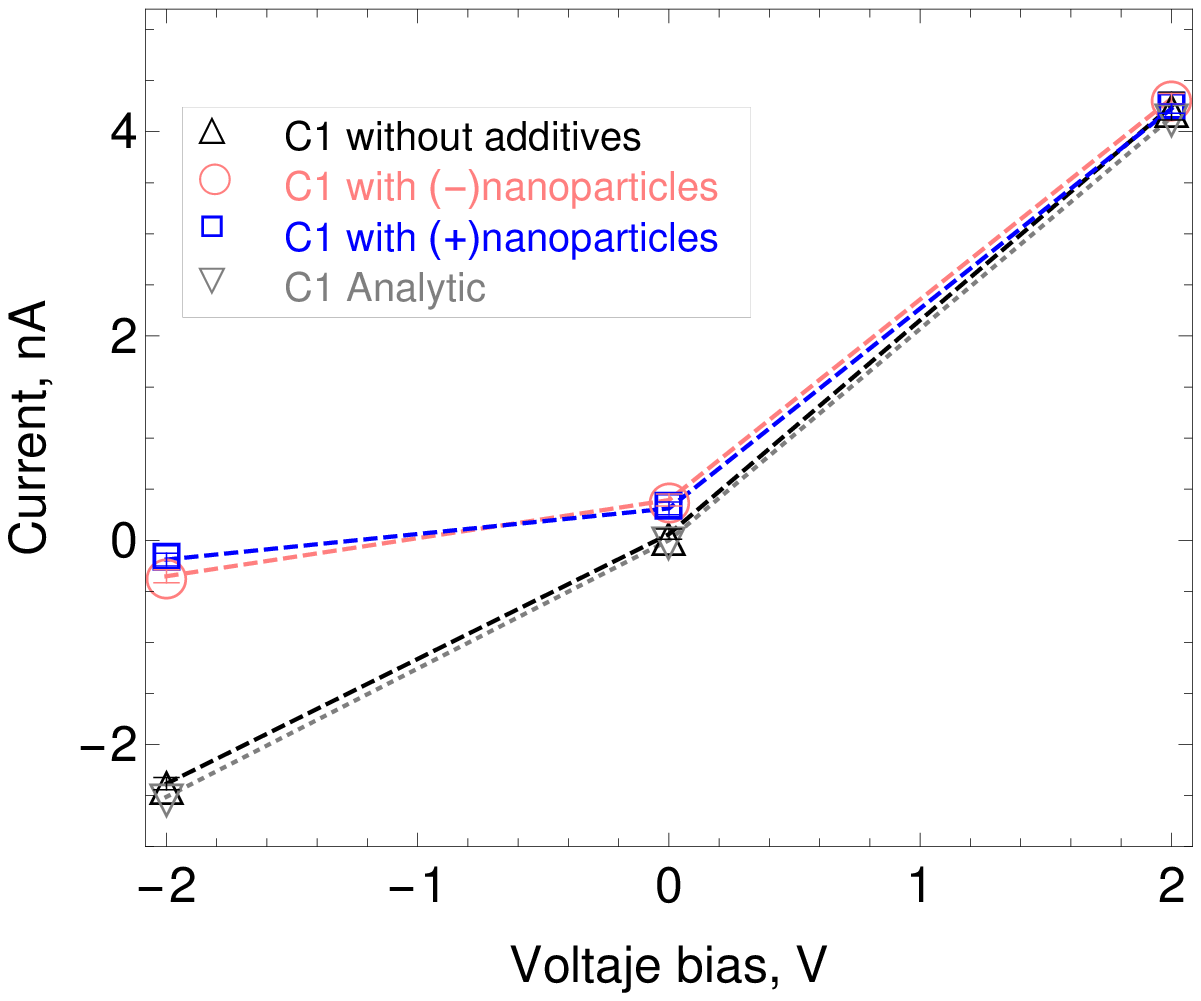}
\put (50,85) {\small A}
\end{overpic}
\hspace{3mm}
\begin{overpic}[width=0.48\linewidth]{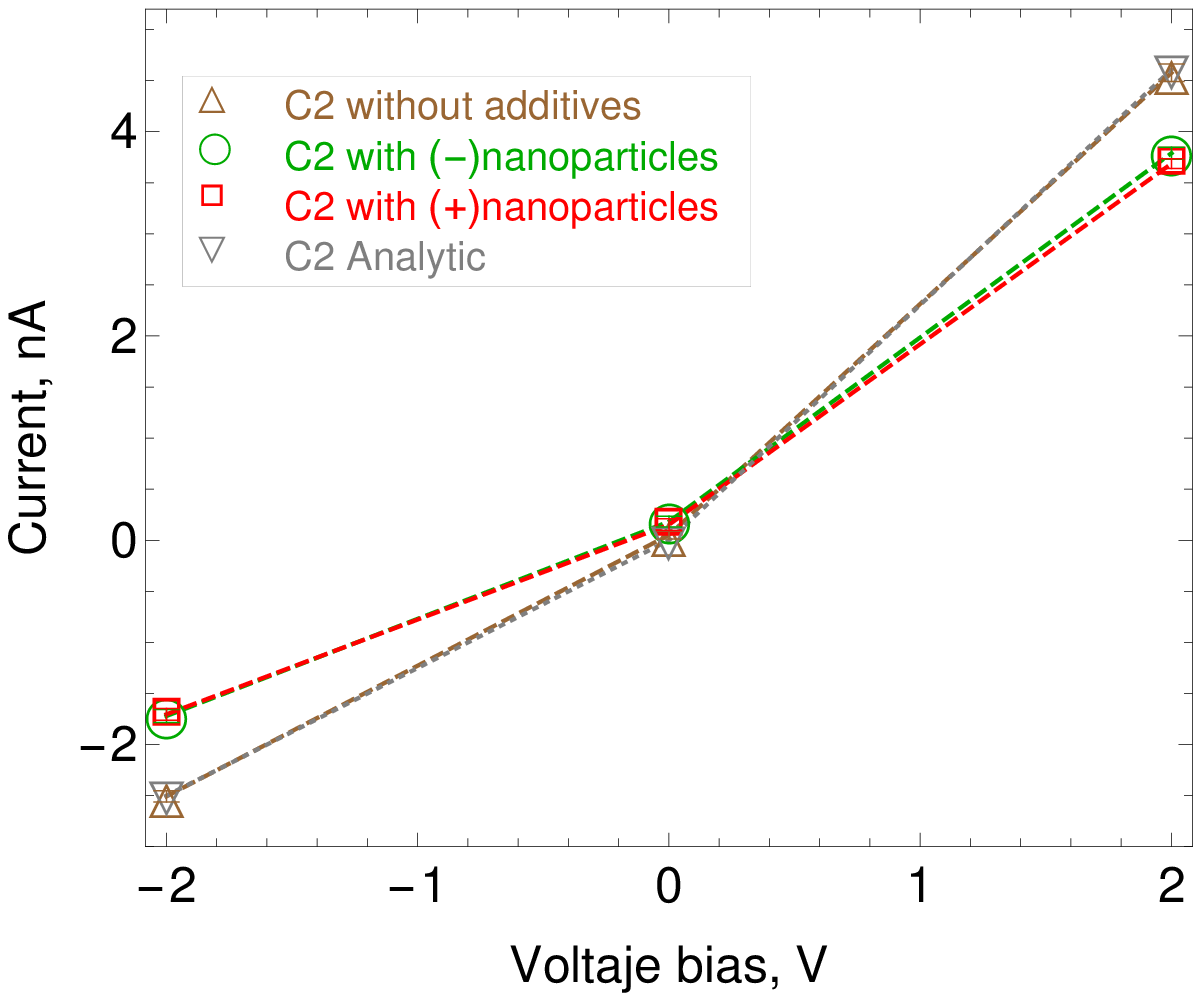}
\put (50,85) {\small B} 
\end{overpic}
\caption{Current-Voltage relationship for the two 
nanochannels considered here with and without nanoparticles.
A) Channel where the two charged sections have the same length, C1 B)
Channel where the negatively charged section is longer than the 
positively charged section, C2. 
\textcolor{black}{The statistical uncertainties for the electric currents 
are calculated using the blocking transformation 
method \cite{flyvbjerg1989error} which is applicable to estimating statistical error in 
averages of time series data by properly accounting for the correlation in the data.}
The bipolar pore model of \citet{green2015asymmetry} is used for the analytical predictions.}
\label{figcurrent}
\end{figure}

\subsection{Electric current rectification}

From a practical perspective, the purpose of computer simulations 
is to design bipolar nanopores that can produce a usable current for a given 
applied voltage bias. Several theoretical and computational works 
have proposed current-voltage relationships for bipolar 
nanopores  \cite{mafe1990model,green2015asymmetry,
moy2000tests, corry2000tests, long2018reverse, 
valisko2019multiscale,fertig2019scaling,fertig2020rectification}.
Figure \ref{figcurrent} shows the
current-voltage relationship for the two slit pores 
considered here, as predicted by our coarse-grained MD calculations.
\textcolor{black}{Details on how the electric current is calculated are given in 
the Methods section below.}
In particular, Figure \ref{figcurrent}A shows the effect of adding 
nanoparticles (positive or negative) 
on the current-voltage relationship for the C1 nanopore. 
It can be observed that, for the $-2~\text{V}$ applied bias,
the electric current is suppressed by the presence of positive and negative charged nanoparticles.
For the  $+2~\text{V}$ bias case, the effect of nanoparticles on the 
current is minor, and falls within the uncertainty of the calculations.
The electric current rectification for C1 with positively charged nanoparticles 
is 22.3, which is about 13 times larger than without nanoparticles. 
The electric current rectification for C1 with negatively charged nanoparticles 
is 12.3, which is about 7 times larger than without nanoparticles. 

Figure \ref{figcurrent}B shows the effect of adding 
nanoparticles, positively or negatively charged 
on the current-voltage relationship for the C2 nanopore. 
First, note that without nanoparticles the rectification ratios in C2, 
$|I_\text{+2V}^\text{C2}|/|I_\text{-2V}^\text{C2}|=1.82$,
and C1 $|I_\text{+2V}^\text{C1}|/|I_\text{-2V}^\text{C1}|=1.78$
are very similar. However, when nanoparticles (either positively or negatively charged) 
are added to C2, the rectification ratio increases only by a factor of about 1.2. 
This represents a much smaller increase than the factor of 13 that is observed in C1
when positively charged nanoparticles are added.
A summary of the rectification factors is given in Table \ref{tab:rect}.

\begin{table}[ht]
\centering
\addtolength{\tabcolsep}{12pt}    
\begin{tabular}{ccc}
\hline
Additive & \multicolumn{2}{c}{Electric current rectification } \\
& \multicolumn{2}{c}{($|I_\text{+2V}|/|I_\text{-2V}|$)}  \\
\hline 
& C1 & C2 \\
\hline
None & $1.78\pm0.05$ & $1.82\pm0.035$  \\ 
\hline
&\multicolumn{2}{c}{Negatively Charged}\\
\hline
10 Particles & $12.3\pm2.05$ & $2.20\pm0.076$ \\
\hline
&\multicolumn{2}{c}{Positively Charged}\\
\hline
10 Particles & $22.3\pm7.3$ & $2.16\pm0.08$ \\
\hline
\end{tabular}
\caption{Summary of the electric current rectification ratios for the different 
systems considered in this work. \textcolor{black}{The uncertainties in these ratios 
are obtained by propagating the error in the electric currents. See the SI 
for the values of the electric currents and uncertainties from which 
the rectification ratios are calculated.} }
\label{tab:rect}
\end{table}

Figure \ref{figcurrent} also shows comparisons 
of the current-voltage relation from the MD simulations used here 
with predictions obtained using the analytic model for bipolar nanopores 
proposed by \citet{green2015asymmetry}.
The analytic model is based on the Poisson-Nernst-Planck (PNP) equations. 
It's main assumptions are local electroneutrality in the reservoirs,
cross-sectional local electroneutrality in the permselective regions 
and equal concentration of the electrolyte in the two reservoirs. 
The analytic predictions of that model have
been verified by solving the fully coupled PNP
equations using the finite elements numerical method.
The analytic expression for the electric current has three distinct terms,
one term accounts for the reservoirs'
Ohmic resistors, which also account for field focusing into
the permselective interfaces. Another term describes the 
electric potential drop over the permselective regions. A third term
describes the Donnan potential jumps at all three interfaces.
The only adjustable parameter of the model is a diffusion coefficient for 
the ions. 

\textcolor{black}{For the predictions shown in Figure \ref{figcurrent} 
the value obtained from fitting the analytic model to the electric currents
obtained in the simulations is
$\mathcal{D}=1.0 \times 10^{-10}~\text{m}^2/\text{s}$. This value can be contrasted
with the diffusion coefficients of the ions inside the nanopores measured in the simulations. 
These are $\mathcal{D}_{\rm Na^+}\approx9.05 \times 10^{-10}~\text{m}^2/\text{s}$ and
$\mathcal{D}_{\rm Cl^-}\approx5.85 \times 10^{-10}~\text{m}^2/\text{s}$. 
See the Methods section below for details on how these diffusion coefficients are calculated. 
This means that to predict similar currents the diffusion coefficient in the analytic model has to be 
about 5 to 9 times smaller than the diffusion coefficients in the MD simulations. 
The difference is reasonable given 
the different levels of description of the two models; the less detailed level of the continuum
analytical model and the more detailed level of the MD simulations.} 
Note That the analytical model is only applicable 
in the cases where there are no nanoparticles 
added. But in those cases the electric current shows agreement with the results obtained 
with the MD simulations employed here.

\textcolor{black}{We have also examined the effect of the nanoparticle's friction 
coefficient on the electric current. This was done by performing additional replicas 
of the simulations but using a friction coefficient for the nanoparticles about 8 times smaller 
than the one used for the results shown in Figure \ref{figcurrent}. See the Methods section below 
for details. Figure S10 and Table S3 
in the SI show comparisons of the electric currents and the rectification ratios obtained 
in the different replicas performed for nanopore C1 with positively 
charged nanoparticles. To within the statistical uncertainty 
of the simulations all the replicas 
give the same electric currents and rectification ratios.}

 \subsection{Mechanism of current rectification}
 
\textcolor{black}{In the simulations 
presented here, the ionic strength of the solutions has been kept
constant when adding the nanoparticles. Therefore one can immediately  rule out 
that the observed increases in current rectification are 
due to an increase or decrease in the concentration of the total charges in the systems.
Also, the concentration of particles has been kept very low, at about $0.7$ mM,
and their mobility is about ten times smaller than the mobility of the ions 
(see the Methods section below). Therefore the contribution of the nanoparticles
to the total electric current flowing through the nanopores is always less than $1\%$
(see Tables S4--S6 in the SI for the individual contributions of each 
charged species to the total electric currents).
Moreover, changing the friction coefficient of the nanoparticles 
by a factor of 8 does not have a significant effect on the observed rectification ratios.
Nevertheless, the nanoparticles can have a significant effect on the charge distribution 
inside the nanopores and therefore can produce significant changes in the nanopore's 
electric potential.}

\begin{figure}[h t]
\center
\includegraphics[width=0.5\linewidth]{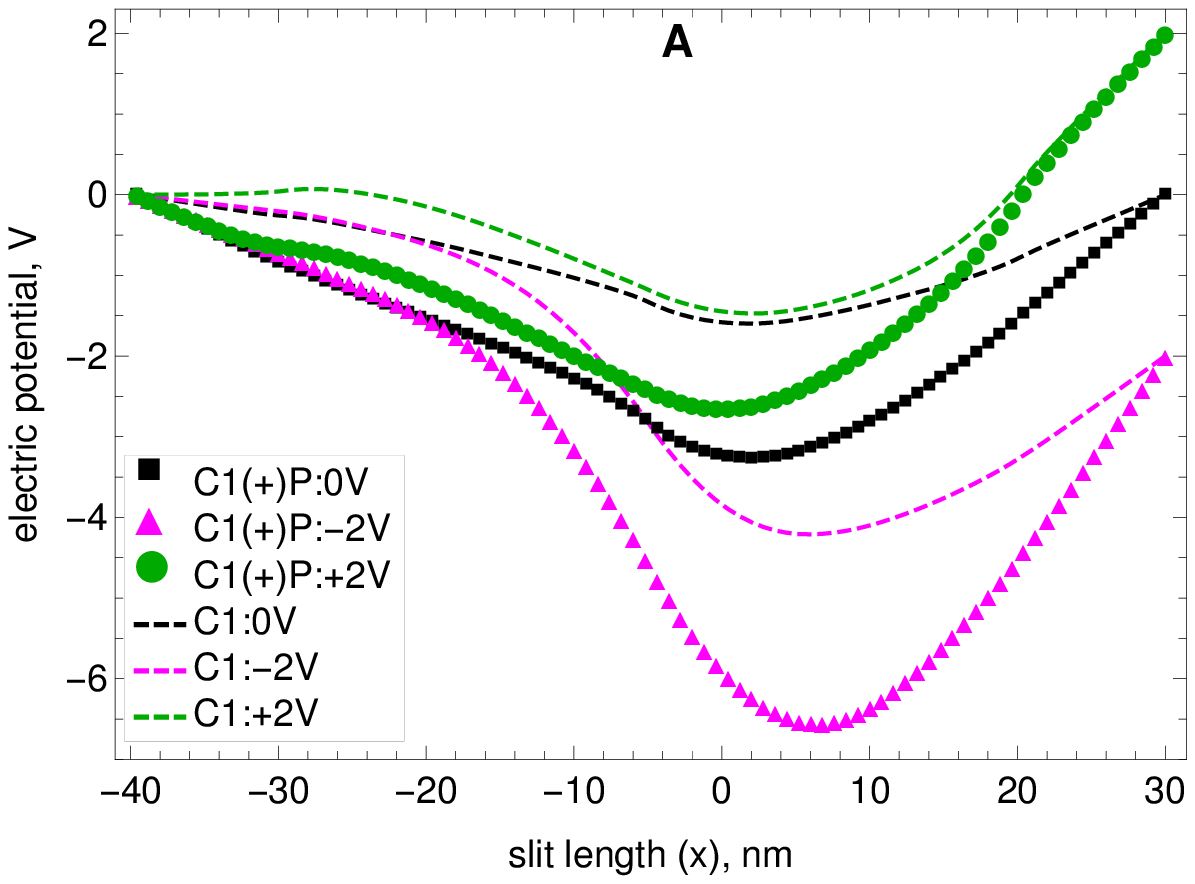}~~
\includegraphics[width=0.5\linewidth]{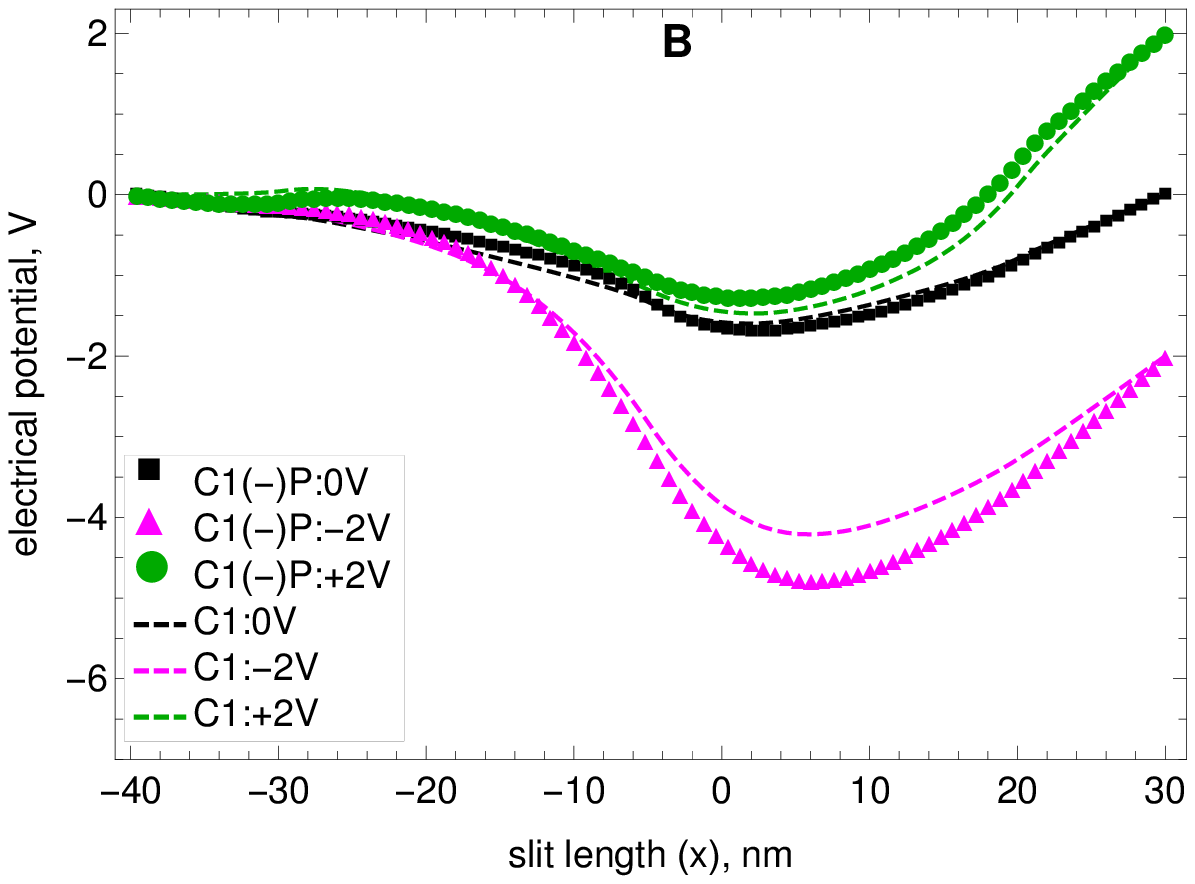}
\caption{Effect on the
electric potential of C1 as a function of channel 
length when A) positively charged nanoparticles are added 
and B) negatively charged nanoparticles are added. 
The electric potential in the channels where nanoparticles are added is shown
with symbols. The electric potentials in the channels without nanoparticles 
are shown with lines. The electric potentials in nanopore C2 are given in the 
Supporting Information file.}
\label{figpotc1}
\end{figure}

\textcolor{black}{Several theoretical studies
based on the PNP equations \cite{mauro1962space, 
coster1965quantitative,sonin1972ion,green2014effect,chinaryan2014effect,
green2015asymmetry,green2015time}, have  shown that}
in a bipolar nanopore, the asymmetric and diode-like behavior
of the current is due to the system layout. 
When the voltage
is applied in a direction that coincides with the inherent
electric field of the nanopore, then the current
is enhanced. When the voltage is applied in the
opposite direction, a so-called reverse bias, then the applied electric field
operates against the inherent electric field, leading to a
reduction in the resultant current \cite{green2015asymmetry}.
The effects of adding positively charged nanoparticles or 
negatively charged nanoparticles on the electric field are shown in Figures \ref{figpotc1} A and B, 
respectively. Results are shown for the electric potential in C1 as a function 
of channel length, $\psi(x)$. \textcolor{black}{The details on how the electric field 
of the nanopores as a function of length, $\psi(x)$, is calculated are given in the Methods 
section below.} Results for C2 are given in Figure S6 of the SI. 

One can clearly observe that the nanoparticles 
have a large effect on the electric potential. 
More specifically, in C1 negatively charged nanoparticles increase the depth of the minimum from
$-1.6~\text{V}$ to $-1.7~\text{V}$, and positively
charged nanoparticles increase the depth of the minimum from
$-1.6~\text{V}$ to $-3.3~\text{V}$.
For the $-2~\text{V}$ applied bias case, negatively charged nanoparticles 
increase the depth of the minimum from $-4.2~\text{V}$ to
$-4.8~\text{V}$. For this negative bias case, positively charged nanoparticles 
increase the depth of the minimum from $-4.2~\text{V}$ to $-6.5~\text{V}$.
For the $+2~\text{V}$ 
applied bias case, negatively charged nanoparticles 
decrease the depth of minimum in the electric potential from $-1.5~\text{V}$ to
$-1.3~\text{V}$. For this positive bias case, positively charged nanoparticles 
increase the depth of the minimum from $-1.5~\text{V}$ to $-2.6~\text{V}$.
\textcolor{black}{In tables S4--S6 of the SI it can be observed that
the deeper minimum in the electric potential observed for $-2~\text{V}$  
bias in the presence of nanoparticles significantly reduces the fluxes of anions
and cations in that case.}
The rectification ratio is increased more by the positively
charged nanoparticles because these produce a much deeper minimum
in the electric potential for the $-2~\text{V}$  bias, thereby causing 
the current to be diminished for that bias while maintaining the 
current enhancement for the $+2~\text{V}$ bias. 

In C2, nanoparticles do not have a significant effect on the 
inherent electric field of the nanopore. Overall, they tend to 
slightly increase the depth of the 
minimum in the electric potential.
This leads to a slight overall reduction in the electric current that is observed 
in that pore upon addition of nanoparticles.

\section{Conclusions}

While recent work has produced significant theoretical and experimental  
progress in our understanding of 
the fundamental physics that govern bipolar nanopores, 
important design challenges still remain with regards to the 
role of nanopore geometry 
and the addition of nanoparticles on current rectification.
In this paper, we have considered two channel geometries,
one in which the two oppositely charged sections have the same length,
and one in which the negatively charged section is about twice the length of the 
negatively charged section. Both of these geometries include
changes in the height as a function of channel length. The cation and anion concentration profiles and 
current-voltage relationship of dilute electrolytes (0.1 M) were determined for neutral, 
positive, and negative applied biases, respectively, and we examined the role of nanoparticles on current rectification.

Our simulations with explicit ions in an implicit solvent have revealed that the current rectification ratio in the 
bipolar nanochannel with a longer negatively charged section is 1.82. 
Negatively charged nanoparticles can raise the rectification to 2.2. 
In a channel with that geometry,
positively charged nanoparticles also increase 
the rectification to 2.2. In the channel where the negatively and positively 
charged sections have the same length, the current rectification without nanoparticles is 1.78. 
However, adding negatively charged nanoparticles increases 
the rectification to 12.3. In this latter channel, positively charged nanoparticles increase the current rectification to 22.3.
These considerable increases in current rectification can be explained by the 
large effect that positively charged nanoparticles have in the inherent electric field of the 
nanopore. Moreover, this effect is strongly dependent on 
the pore geometry, its charge distribution and the sign of the nanoparticles' charge. These findings could have significant implications for design of improved current rectification systems based on nanopore diodes, and we hope that they will stimulate experimental work. 

\section{Methods}

\subsection{Ions model}

The coarse-grained, CG, parameters of the ion-ion effective potentials
were obtained using relative entropy coarse-graining, RE-CG, \cite{shell2008relative}, 
which provides a systematic approach to obtaining effective potentials for use in
CG simulations, given a mapping function from all-atom, AA, 
to CG coordinates \cite{hinckley2014ions}.
The CG non-bonded effective potentials between ions are
represented as the sum of a Coulombic interaction and a correction
term, $U_{\rm corr}$, as follows,
\begin{eqnarray}
U_{\rm ion-ion}(r_{ij})=\frac{q_iq_j}{4\pi\epsilon_0\epsilon(T)r_{ij}}+U_{\rm corr}(r_{ij}).
\end{eqnarray}
Here $q_i$ and $q_j$ are the charges of the $i^{\rm th}$ and $j^{\rm th}$ ion.
The dielectric permittivity of vacuum is given by $\epsilon_0$. Here 
$\epsilon(T)=249.4-0.788T/{\rm K}+7.20\times10^{-4}(T/{\rm K})^2$ is the solution
dielectric, $T$ is the absolute temperature
and $r_{ij}$ is the inter-ion separation. $U_{\rm corr}$ is represented using
cubic splines, and corrects the Coulombic potential to
account for the effects of hydration and stearic overlap. The use
of cubic splines permits the reproduction of subtle effects that
could not be easily resolved with analytical expressions.
The effective potentials recreate essentially all features of
the radial distribution functions from AA simulations \cite{hinckley2014ions}.
The effective soft-core repulsion  (\textit{i.e.} radius) that
arises after systematic coarse-graining is $0.2494~\text{nm}$ for
Sodium (Na$^{+}$) ions and $0.4478~\text{nm}$ for Chlorine (Cl$^{-}$) ions.

\subsection{Nanopres model}

The nanopores considered here are channels  with a slit geometry 
that is periodic in the $x$ and $y$ coordinates. 
The total size of the simulation box is set to
$70~\text{nm}\times~20~\text{nm}\times30.1~\text{nm}$.
The height of the channels and the wall charge density
as a function of length, $\left[h(x),\sigma(x)\right]$,  is given by,
\begin{align}\label{geo}
\left[h(x), \sigma(x)\right]=\left\{
\begin{array}{ccccc}
\left[30~\text{nm},~~0\right] & & & & -40~\text{nm} \leq x  \leq -30~\text{nm} \\
\left[17~\text{nm},~~\sigma_1\right] & & & & -30~\text{nm} \leq x  \leq -\ell \\
\left[10~\text{nm},~~ \sigma_2\right] & & & & -\ell \leq x \leq 20~\text{nm} \\
\left[30~\text{nm},~~0\right] & & & & 20~\text{nm}  \leq x  \leq 30~\text{nm}.
\end{array}
\right.
\end{align}
Two types of bipolar nanochannels were simulated.
One in which the two oppositely charged sections have the same length
and another one in which the positively charged section is about four times shorter than
the negatively charged section.
The first channel is referred to as C1. In it
$\ell=5~\text{nm}$ in eq. (\ref{geo}). In this channel
for $-30~{\rm nm }\leq x\leq  -5~{\rm nm }$ the wall charge density is
$\sigma_1=+0.1~C/{\rm m}^2$ and
the surface area of the  wall in this section is $S_1=1000~\text{nm}^2$. 
For $-5~{\rm nm }\leq x\leq  20~{\rm nm }$ the wall charge density is
$\sigma_2=-0.1~C/{\rm m}^2$, and the surface area of the  wall in this section is
also $S_2=1000~\text{nm}^2$. The wall charge density is 
$\sigma_{\neq 1,2}=0~C/{\rm m}^2$ for the rest of the walls, 
including the walls between sections of different 
height and walls that are perpendicular to the $x$ axis.
The sections at the end of the channel, where 
the wall charge is zero, are intended to act as reservoirs.
The total volume of C1 is $V_{C1}=25500~\text{nm}^3$.

The second channel is denoted C2 and in it the
negatively charged section is four times longer than the 
positively charged section. Therefore $\ell=20~\text{nm}$ in eq. (\ref{geo}). And
for $-30~{\rm nm }\leq  x\leq  -20~{\rm nm }$ the wall charge density is
$\sigma_1=+0.24~C/{\rm m}^2$ and
the surface area of the  wall in this section is $S_1=400~\text{nm}^2$. 
For $-20~{\rm nm }\leq  x\leq  20~{\rm nm }$ the wall charge density is 
$\sigma_2=-0.08~C/{\rm m}^2$, and the surface area of the  wall in this section is
$S_2=1600~\text{nm}^2$. The wall charge density is 
$\sigma_{\neq 1,2}=0~C/{\rm m}^2$ for the rest of the walls, 
including the walls between sections of different 
height and walls that are perpendicular to the $x$ axis.
The sections at the end of the channel, where 
the wall charge is zero, are intended to act as reservoirs.
The total volume of C2 is $V_{C2}=23400~\text{nm}^3$.

The systems are always globally neutral, 
\textit{i.e.}, $S_1\sigma_1+S_2\sigma_2=z_c e n_c+z_a e n_a+z_p e n_p$.
Where $n_a$ is the total number
of anions (chloride) in the simulation box and $e z_a =-e$ is the charge of each anion. 
$e\approx1.6\times10^{-19}~\text{C}$ is the elementary charge.
$n_c$ is the total number of cations
(Sodium) and  $e z_c=+e$ is the charge of each cation.
$n_p$ is the total number of nanoparticles in the simulation box and 
$e z_p$ is the charge of each nanoparticle. \textcolor{black}{
The Debye length which is used to estimate the size of the electrical double layer is
$\lambda_{\rm D}=\sqrt{\dfrac{\epsilon_0 \epsilon V k_{\rm B} T}{e^2 \sum_{i=1}^N z_i^2}}$.
Where $V$ is the volume of the pore and $N$
is the total number of charged particles in the solution that flows 
through the pore. In all the simulations presented in this
work $\lambda_{\rm D}\approx0.95~\text{nm}$.}
 
In our coarse grained MD simulations the channel walls are modeled 
with a combination of neutral and charged sites that are kept fixed in space
at their initial position. The charged wall sites are placed at random 
positions along the walls. 
The proportion of charged to neutral sites in a (charged) 
wall is determined by the wall charge density, 
the surface area of the wall, and by the charge of the wall charged sites.
Each charged wall site is assigned a charge of 
$\pm 0.25 e$ and interacts with the ions 
through Coulomb interactions.

Additionally, all wall sites, charged and uncharged,
interact with the ions in solution through a repulsive Lenard-Jones potential,
\begin{align}\label{rlj_pot}
U_{\rm rep}(r_{ij})=
\left\{
\begin{array}{cc}
4 \epsilon \left[ \left( \dfrac{\sigma}{r_{ij}} \right)^{12}-
\left( \dfrac{\sigma}{r_{ij}} \right)^{6}\right] & r_{i,j} < r_{\rm cut} \\
0 &  r_{\rm cut} \geq r_{\rm cut}.
\end{array}
\right.
\end{align}
where $\epsilon = 1.68~k_{\rm B}T$, $\sigma = 0.4$ nm.
A cutoff of $r_{\rm cut}=2^{1/6}\sigma$ is applied, and  $r_{ij}$ is the 
distance between the ion and the wall site. 

\textcolor{black}{To apply the electric potential 
bias between the two ends of the bipolar nanopores
we use a method commonly used in MD simulations for this purpose 
\cite{gumbart2012constant}.
The method consists of applying a uniform electric field throughout the 
entire simulated periodic cell containing the nanopore.
The external electric field $\bm{E}(\bm{r})$ is related to the
external electric potential, $\psi_{E}(\bm{r})$,
by $\bm{E}(\bm{r})=-\nabla \psi_{E}(\bm{r})$.
In our simulations the external
electric field is applied in the $x$ direction, $\bm{E}(\bm{r})=E_x\bm{\delta}_x$, 
where $\bm{\delta}_x$ is the unit vector in the $x$ direction and 
$E_x=\pm0.0286~\text{V}/\text{nm}$.
The total length of the pores in the 
$x$ direction is $L=70~\text{nm}$. 
The voltage bias is defined as $\Delta V=\psi_{E}(x=30~\text{nm})-\psi_{E}(x=-40~\text{nm})$.
Therefore when $E_x=-0.0286~\text{V}/\text{nm}$,
$\Delta V=-E_x/L=+2~\text{V}$.
And when $E_x=+0.0286~\text{V}/\text{nm}$,
$\Delta V= -E_x/L=-2~\text{V}$.}

\subsection{Dynamics}

The coarse grained MD simulations were performed using
LAMMPS \cite{plimpton1995fast} in the NVT
ensemble using the Gronbench-Jensen Farago (G-JF) Langevin thermostat
\cite{gronbech2013simple, gronbech2014application}, with a damping factor of
$\tau_{\rm m}=10 \times 10^{-15}~\text{s}$.
The friction coefficient of ion $i$, $\zeta_i$, is related to the
damping factor $\tau_{\rm m}$ by $\zeta_i=\dfrac{m_i}{\tau_{\rm m}}$, 
where $m_i$ is the mass of each ion. 
For Sodium $m_{{\rm Na}^+}=3.82\times10^{-23}~\text{g}$
(\textit{i.e.} $22.9898~\text{g/mol}$) and for 
Chlorine $m_{{\rm Cl}^-}=5.889\times10^{-23}~\text{g}$
(\textit{i.e.} $35.453~\text{g/mol}$).
Therefore $\zeta_{{\rm Na}^+}=3.82\times10^{-9}~\text{N s}/\text{m}$
and $\zeta_{{\rm Cl}^-}=5.89\times10^{-9}~\text{N s}/\text{m}$.
A time step $\Delta t = 2\times10^{-15}~\text{s}$ 
and a temperature of $297~\text{K}$ were used in all simulations.
We relied on the particle mesh Ewald
solver to calculate electrostatic interactions, with a real space cutoff of 1.2 nm.
We used the modification of the three-dimensional 
Ewald summation technique for calculation of
long-range Coulombic forces in systems 
with a slab geometry that are periodic in two dimensions
and have a finite length in the third dimension\cite{yeh1999ewald}.

\subsection{Nanoparticle model}

Each nanoparticle is modeled by placing eight charged sites in the corners of a cube 
with sides of length  $0.8~\text{nm}$. 
Each site in the nanoparticle is assigned a charge of $\pm e$, therefore 
the total charge of each nanoparticle is $ez_p=-8e$. In our simulations, 
the relative distance between these sites is held approximately constant by 
using stiff harmonic springs. More specifically, adjacent sites are bonded by a spring with 
an equilibrium length of $0.8~\text{nm}$ and a spring constant of 
$1.2\times10^5~k_{\rm B}T/{\rm nm}^2$. 
Sites in opposite corners of the cube are bonded by a harmonic potential with equilibrium length 
of $1.39~{\rm nm}$ and an energy barrier of $1.2\times10^5~k_{\rm B}T/{\rm nm}^2$.
Triplets of adjacent sites in the nanoparticles are also constrained with an angle 
potential with equilibrium value of $90^\circ$ and an energy barrier 
of $5.1\times10^2~k_{\rm B}T$. \textcolor{black}{
To estimate the charge density of the nanoparticles we
consider a cube of side length $0.8$ nm.  
Therefore each nanoparticle has a surface area of $3.84~\text{nm}^2$.
Since each nanoparticle has eight charged sites with a charge of $\pm e$ 
the surface charge density is $\pm0.33~\text{C}/\text{m}^2$.}
The total mass of each nanoparticle is 
$m_{p}=305.6\times10^{-23}~\text{g}$ (\textit{i.e.} $1839~\text{g/mol}$). The same damping 
factor used for the ions is employed for the nanoparticles in the Langevin thermostat.
Therefore the friction coefficient 
of each nanoparticles is $\zeta_{p}=305.6\times10^{-9}~\text{N s}/\text{m}$.
The nanoparticle sites as well as the charged wall sites interact with the ions 
and with each other through 
Coulomb's law. All wall sites, charged and uncharged,
as well as the nanoparticle sites also
interact with the ions and with each other
through the repulsive Lenard-Jones potential
given in eq. \ref{rlj_pot}. 

\subsection{Diffusion coefficients}

\textcolor{black}{The diffusion constant of each ionic species in the bulk can be 
calculated \textit{a priori} from the friction coefficients specified above using 
the Einstein relation, \textit{i.e.} $\mathcal{D}=k_{\rm B} T/\zeta$. The values that 
result from that calculation are,
$\mathcal{D}_{\rm Na^+}^{\rm bulk}\approx1.1 \times 10^{-12}~\text{m}^2/\text{s}$,
$\mathcal{D}_{\rm Cl^-}^{\rm bulk}\approx6.9 \times 10^{-13}~\text{m}^2/\text{s}$ and
$\mathcal{D}_{p}^{\rm bulk}\approx1.3 \times 10^{-14}~\text{m}^2/\text{s}$.
However the diffusion coefficients of charged species
inside charged nanopores can be significantly different from diffusion coefficients in the
bulk \cite{valisko2019multiscale}. Therefore we calculate the  diffusion coefficient of 
anions, cations and nanoparticles inside the confined environment of the nanopores from 
the mean-squared displacement, $\langle \Delta r ^2(t) \rangle_{\rm eq}$, 
of each species inside the nanopores. 
This calculation is done without a voltage bias applied and therefore
$\langle ... \rangle_{\rm eq}$ indicates an average at equilibrium. 
Specifically the diffusion constants are obtained from the relation
$\mathcal{D}=\langle \Delta r ^2(t) \rangle_{\rm eq}/(6 t)$. See
Figure S7 of the SI for plots of the mean-squared displacements and 
see Table S2 also in the SI 
with the diffusion coefficients for each species in each system. The 
diffusion coefficient for each species does not change significantly from one system
to another. The average values are, 
$\mathcal{D}_{\rm Na^+}^{\rm eff}\approx9.05 \times 10^{-10}~\text{m}^2/\text{s}$,
$\mathcal{D}_{\rm Cl^-}^{\rm eff}\approx5.85 \times 10^{-10}~\text{m}^2/\text{s}$ and
$\mathcal{D}_{p}^{\rm eff}\approx0.24 \times 10^{-10}~\text{m}^2/\text{s}$.}

\textcolor{black}{The effective diffusion 
coefficients inside the nanopores observed in the simulations can be 
compared with the values obtained from a simple hydrodynamic calculation. 
The simple hydrodynamic calculation assumes that ions and noanoparticles 
can be treated as spheres.
The diffusion coefficients are calculated from the relation 
$\mathcal{D}=k_{\rm B} T/\zeta$, with $\zeta=6 \pi R \eta$ where $\eta$ 
is the viscosity of the solvent and $R$ is the hydrodynamic radius of each species. For the
nanoparticles we consider that the eight sites that compose it are at a distance 
$0.69$ nm from its center. Also since each site interacts with other species through 
eq. (\ref{rlj_pot}) we add $0.4\times2^{1/6}=0.44$ nm to the previous value. 
Therefore $R_p\approx1.14$ nm. The soft-core repulsion radii given above 
in the Ions model subsection are used for the ions. 
Then, using $\eta\approx1$ mPas, the diffusion coefficients given by this calculation are 
$\mathcal{D}_{\rm Na^+}^{\rm hyd}\approx8.7 \times 10^{-10}~\text{m}^2/\text{s}$ and
$\mathcal{D}_{\rm Cl^-}^{\rm hyd}\approx4.8 \times 10^{-10}~\text{m}^2/\text{s}$ and
$\mathcal{D}_{p}^{\rm hyd}\approx1.9 \times 10^{-10}~\text{m}^2/\text{s}$. 
The effective diffusion coefficients for the ions inside the nanopore agree well with the 
simple hydrodynamic calculation. For the nanoparticles the effective diffusivity
in the simulations is about 7.9 times smaller than the one obtained in the simple hydrodynamic
calculation. Moreover, below we show that the current rectification results can also
be replicated when a smaller friction coefficient is used for the nanoparticles in the simulations.}

\subsection{Electric current}

\textcolor{black}{The electric current that flows trough the nanopore as a function of time, 
is calculated using a commonly 
employed method \cite{crozier2001model, aksimentiev2005imaging,di2021geometrically},
$I(t)=-\displaystyle \sum_{j=1}^N \dfrac{\Delta x_j(t) ez_j}{L\Delta t}$,
where the sum is over all the ions and nanoparticles, $N$, and $ez_j$ is the charge 
of each species. $\Delta x_j(t)$ is the displacement 
in the $x$ direction of charged species $j$ on the time interval $\Delta t$ and $L$ is the 
total length of the channel. Plots 
of $I(t)$ during $24~\text{ns}$ of our production runs are given in Figure S5 of the SI
for all the different systems considered in here. Moreover, 
in Figure \ref{figcurrent} we have reported the 
time average of $I(t)$ over those $24~\text{ns}$ of our production runs for all 
the systems considered in this work. The statistical uncertainties
for the electric currents plotted in Figure \ref{figcurrent}
and also given in Table S1 in the SI were calculated using the blocking transformation 
method of Flyvbjerg and Petersen
\cite{flyvbjerg1989error}, specifically Eq. (28) in that reference. 
The blocking transformation  method is applicable to estimating statistical error in 
averages of time series data and properly accounts for the correlation in the data.}

\subsection{Electric potential}

\textcolor{black}{To calculate the total electric potential of the nanopores as a 
function of length, $\psi(x)$, 
shown in Figure (\ref{figpotc1}) we use the Poisson equation integrated over the $y$ and 
$z$ coordinates, \textit{i.e.}
$\dfrac{\psi^2(x)}{dx^2}=-\dfrac{\rho_e(x)}{\epsilon_0 \epsilon}$. Where 
$\rho_e(x)=N_{\rm A} e \left[z_c c_c(x)+z_ac_a(x)+z_pc_p(x)\right]+\dfrac{\sigma(x)}{h(x)}$ 
 is the charge density as a function of the channel's length 
(\textit{i.e.} averaged over the channel's width and depth). 
Where $N_{\rm A}$ is Avogadro's number, $\sigma(x)$ and $h(x)$ were defined in eq. (\ref{geo}).
$c_c(x), c_a(x)$ and  $c_p(x)$ are molar concentration profiles of cations, 
anions and nanoparticles respectively.
Each concentration profile is obtained by averaging $10^6$ simulation frames taken during
$24~\text{ns}$ of simulation after the systems have reached steady-state. In each frame
the simulation box is divided into cubic bins of $(0.8 \times 20 \times 0.2)~\text{nm}^3$. 
The number of particles of each species inside each bin are counted
to calculate the concentrations. The steady-state molar 
concentration profiles are given in Figure \ref{figconcpp} and Figures S3 and S4 in the SI.
The Poisson equation is solved with boundary conditions, $\psi(x=-40~\text{nm})=0$ and 
$\psi(x=30~\text{nm})=-E_x/L$.}

\subsection{Reproducibility}

\textcolor{black}{To check for reproducibility of the rectification results presented here we have 
performed additional replicas of the simulations for the system that shows the 
largest rectification ratio. Moreover we have also performed those replicas using a 
friction coefficient for the nanoparticles about 8 times smaller than the
specified in the Nanoparticel model subsection above. More precisely, a friction coefficient
of $\zeta_{p}=38.7\times10^{-9}~\text{N s}/\text{m}$ was used for the 
nanoparticles in the additional replicas. The friction coefficients of the ions were kept the same 
as those given in the Dynamics subsection above.
Snapshots of the othographic projections of the initial conditions and steady-state 
configurations of these replicas are shown in Figure S8 of the SI. The
mean-squared displacement at equilibrium, $\langle \Delta r ^2(t) \rangle_{\rm eq}$, 
of each species inside the nanopores for each replica are given in Figure S9. 
The average values of the diffusion coefficients obtained from these 
$\langle \Delta r ^2(t) \rangle_{\rm eq}$ are, 
$\mathcal{D}_{\rm Na^+}^{\rm eff}\approx9.1 \times 10^{-10}~\text{m}^2/\text{s}$,
$\mathcal{D}_{\rm Cl^-}^{\rm eff}\approx6.1 \times 10^{-10}~\text{m}^2/\text{s}$ and
$\mathcal{D}_{p}^{\rm eff}\approx1.0 \times 10^{-10}~\text{m}^2/\text{s}$. Note 
that for the nanoparticles this diffusion coefficient is about 4 times larger than the one 
obtained in the simulations with the larger friction coefficient. Figure S10 and Table S3 
in the SI show comparisons of the electric currents and the rectification ratios obtained 
in the different replicas performed.}

\section{Acknowledgment}

\noindent This work is supported by the Department of Energy, 
Basic Energy Sciences, through the AMEWS EFRC Center. 
The development of advanced sampling codes is supported by the Department of Energy, 
Basic Energy Sciences, through MiCCoM.

\section{Supporting Information}

\noindent \textcolor{black}{In the SI we provide the 
following additional data and figures: Versions of Figures \ref{figRP} and 
\ref{figconcpp} when the particles are negatively 
charged; the version of Figure \ref{figpotc1} for pore C2;
figures with the concentration of the ions as a function of the channel height in the 
positively and negatively charged regions of the nanopores; figures 
with the concentration of nanoparticles as a function of the channel length; 
plots of the electric current as a 
function of time; a table with the averages and uncertainties of the  electric
currents; mean-squared displacements of ions and nanoparticles in the
nanopores at equilibrium;  a table with the diffusion coefficients of ions and nanoparticles 
in the different nanopores considered here.}

\section{Notes}

\noindent The authors declare no competing interests


\providecommand{\latin}[1]{#1}
\makeatletter
\providecommand{\doi}
  {\begingroup\let\do\@makeother\dospecials
  \catcode`\{=1 \catcode`\}=2 \doi@aux}
\providecommand{\doi@aux}[1]{\endgroup\texttt{#1}}
\makeatother
\providecommand*\mcitethebibliography{\thebibliography}
\csname @ifundefined\endcsname{endmcitethebibliography}
  {\let\endmcitethebibliography\endthebibliography}{}

\end{document}